\documentclass[final,superscriptaddress,pra,twocolumn,10pt,aps,showpacs,showkeys]{revtex4-1}

\usepackage[separate-uncertainty=true]{siunitx}
\usepackage{color,xcolor}
\usepackage{mdframed}
\usepackage{amsmath}
\usepackage{float}
\usepackage{graphicx}

\makeatletter
\def\verbatim{\small\@verbatim \frenchspacing\@vobeyspaces \@xverbatim}
\makeatother

\begin{document}

\title{A Case Study: Novel Group Interactions through Introductory Computational Physics}

\keywords{Physics Education Research, Computation, Debugging, Case Study}
\pacs{01.40.Fk, 01.50.H-}

\author{Michael J. \surname{Obsniuk}}
\affiliation{Department of Physics and Astronomy, Michigan State University, East Lansing, MI 48824}
\author{Paul W. \surname{Irving}}
\affiliation{Department of Physics and Astronomy, Michigan State University, East Lansing, MI 48824}
   \affiliation{CREATE for STEM Institute, Michigan State University, East Lansing, MI 48824}
\author{Marcos D. \surname{Caballero}}
   \affiliation{Department of Physics and Astronomy, Michigan State University, East Lansing, MI 48824}
   \affiliation{CREATE for STEM Institute, Michigan State University, East Lansing, MI 48824}

\begin{abstract}
With the advent of high-level programming languages capable of quickly rendering three-dimensional simulations, the inclusion of computers as a learning tool in the classroom has become more prevalent.  Although work has begun to study the patterns seen in implementing and assessing computation in introductory physics, more insight is needed to understand the observed effects of blending computation with physics in a group setting.  In a newly adopted format of introductory calculus-based mechanics, called Projects and Practices in Physics, groups of students work on short modeling projects -- which make use of a novel inquiry-based approach -- to develop their understanding of both physics content and practice.  Preliminary analyses of observational data of groups engaging with computation, coupled with synchronized computer screencast, has revealed a unique group interaction afforded by the practices specific to computational physics -- problem debugging.
\end{abstract}

\maketitle

\section{Introduction}

Since the development of reasonably affordable and fast computers, educators have argued for their inclusion in the classroom as both a learning aid and a tool \cite{DiSess1986,Shecker1993}.  With the recent development of high-level programming languages capable of quickly rendering three-dimensional real-world simulations, the argument for the inclusion of computers in the classroom has not only persisted, but has grown \cite{Chonacky2008}.  Well into the 21$^{\rm st}$ century, with its prevalence in modern physics research, we see computation being increasingly referred to as the ``third pillar'' of physics -- along with the more canonical pillars of theoretical and experimental physics \cite{Chabay2008}.

One physics curriculum that includes a computational element is Matter \& Interactions (M\&I).  The M\&I curriculum differs from a traditional one not only through the inclusion of computation (VPython), but also through its emphasis on fundamental physics principles and the addition of a microscopic view of matter \cite{Chabay2004,Chabay1999}.  Recent work \cite{Kohlmyer2005,Caballero2012} involving students' use of VPython with the M\&I curriculum has begun to analyze the patterns seen in implementing and assessing the use of computation in introductory physics courses.

However, numerous questions remain unanswered regarding the processes observed while groups of students work together to model real world phenomena computationally.  We extend our research to a novel implementation of M\&I with an emphasis on computation in a group setting, called Projects and Practices in Physics (${\rm P}^{3}$), where students negotiate meaning in small groups, develop a shared vision for their group's approach, and employ science practices to navigate complex physics problems successfully.  Borrowing from the rich literature of computer science education research \cite{McCauley2008,Murphy2008}, we use the notion of computational debugging in a physics context to help uncover the salient practices unique to computational physics problems.

In this paper, we present a case study of a group of students immersed in this ${\rm P}^{3}$ environment solving a computational problem.  This problem requires the translation of a number of fundamental physics principles into computer code and vice versa.  Our analysis consists of qualitative observations in an attempt to describe, rather than generalize, the computational interactions, debugging strategies, and learning opportunities bolstered by this novel environment.  We have observed two distinct strategies well suited to computational tasks, demonstrating the benefit of the inclusion of this modern tool.

\section{Assumptions}

In an increasingly technological world, we have at our disposal computers well suited for the most procedural and tedious, yet indispensable of STEM tasks.  Accordingly, in ${\rm P}^{3}$, we treat the computer as an indispensable \emph{modern tool} with which students must familiarize themselves.  In spite of the fact that modern computers \emph{take} meaningful direction quite well, they do not yet posses the faculty to \emph{generate} meaningful direction on their own.  This necessitates a human factor, in which groups of students must leverage their understanding of fundamental physical principles to model real world phenomena computationally (i.e., generate meaningful direction to a computer).

\begin{figure}
\resizebox{0.9\columnwidth}{!}{\includegraphics{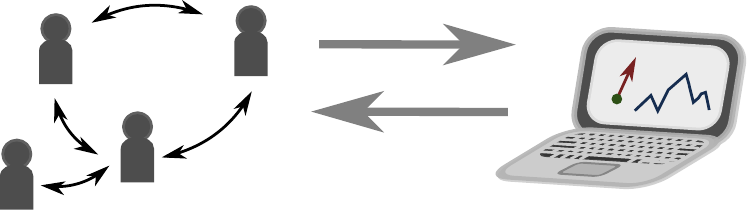}}
\caption{A group of students interacting with VPython where social exchanges focus on FPPs and desirable strategies are observed through discussion and action.}
\label{interactions}
\end{figure}

We focus our research on the social exchanges between group members and the interactions between group and computer, illustrated in Fig.~\ref{interactions}, to begin to understand the ways in which computation can influence learning.  Particularly, we are interested in the interactions occurring simultaneously with social exchanges of fundamental physics principles (FPPs) specific to the present task (e.g., discussing $d\mathbf{r}=\mathbf{v}\,dt$ on a motion task) and the display of desirable strategies (e.g., divide-and-conquer).  These group-computer interactions vary in form, from actively sifting through lines of code, to observing a three-dimensional visual display.

One previously defined computational interaction that reinforces desirable strategies \cite{Murphy2008}, borrowing from computer science education research, is the process of debugging.  Computer science \cite{McCauley2008} defines debugging as a process that comes after testing \emph{syntactically} correct code where programmers ``find out exactly where the error is and how to fix it.''  Given the generic nature of the application of computation in computer science environments (e.g., data sorting, poker statistics, or ``Hello, World!'' tasks), we expect to see different strategies more aligned with the goals of a computational \emph{physics} environment.  Thus, we extend this notion of computer science debugging into a physics context to help uncover the strategies employed while groups of students debug \emph{fundamentally} correct code that produces unexpected physical results.

\section{Data}

In Fall 2014, ${\rm P}^{3}$ was run at Michigan State University in the Physics Department.  It was this first semester where we collected \emph{in situ} data using three sets of video camera, microphone, and laptop with screencasting software to document three different groups each week.  From the subset of this data containing computational problems, we \emph{purposefully sampled} a particularly interesting group in terms of their computational interactions, as identified by their instructor.  That is, we chose our case study not based on generalizability, but rather on the group's receptive and engaging nature with the project as an \emph{extreme case} \cite{Flyvbjerg2006}.

The project that the selected group worked on for this study consists of creating a computational model to simulate the geostationary orbit of a satellite around Earth.  In order to generate a simulation that produced the desired output, the group had to incorporate a position dependent Newtonian gravitational force and the update of momentum, using realistic numerical values.  The appropriate numerical values are Googleable, though instructors encouraged groups to solve for them analytically.

This study focuses on one group in the fourth week of class (the fourth computational problem seen) consisting of four individuals: Students A, B, C, and D.  The group had primary interaction with one assigned instructor.  Broadly, we see a 50/50 split on gender, with one ESL international student.  Student A had the most programming experience out of the group.  It is through the audiovisual and screencast documentation of this group's interaction with each other and with the technology available that we began our analysis.  The majority of the inferences made are based on the audio of discussion, while supporting evidence is gathered from the actions (e.g., writing equations or gesturing) observed through video.

\section{Analysis}

To focus in on the group's successful physics debugging occurring over the $\SI{2}{\hour}$ class period, we needed to identify phases in time when the group had recognized and resolved a physics bug.  These two phases in time, \emph{bug recognition} and \emph{bug resolution} are the necessary limits on either side of the process of \emph{physics debugging}.  We identified these two bounding phases at around $\SI{60(5)}{\minute}$ into the problem, and further examined the process of debugging in-between.  That is, we focused on the crucial moments surrounding the final modifications that took the code from producing unexpected output to expected output.


\subsection{Bug recognition}

At around $\SI{55}{\minute}$ into the problem, following an intervention from their instructor, the group began to indicate that they were at an impasse:  \begin{verbatim}
    SB:  We're stuck.
    SD:  Yeah...
\end{verbatim}  The simulation clearly displayed the trajectory of the satellite falling into the Earth -- not the geostationary orbit they expected -- as observed on the screencast.  This impasse was matched with an indication that they believed the FPPs necessary to model this real world phenomenon were incorporated successfully into the code:  \begin{verbatim}
    SB:  And it's gonna be something really 
         dumb too.
    SA:  That's the thing like, I don't think 
         it's a problem with our understanding 
         of physics, it's a problem with our 
         understanding of Python.
\end{verbatim}  Instead of attributing the unexpected output with a mistake in their understanding or encoding of FPPs, they instead seemed to place blame on the computational aspect of the task.

During this initial phase, we see a clear indication that the group has recognized a bug -- there is an unidentified error in the code, which must be found and fixed:  \begin{verbatim}
    SA:  I don't know what needs to change 
         here...
    SD:  I mean, that means we could have 
         like anything wrong really.
\end{verbatim}  Although they have identified the existence of the bug, they still are not sure how to fix it -- this necessitates the process of debugging.

\subsection{Physics debugging}

Within the previously identified phase of bug recognition, the group developed a clear and primary task: figure out exactly how to remove the bug.  Eventually, following a little off-topic discussion, the group accepted that in order to produce a simulation that generates the correct output, they must once again delve into the code to check every line:  \begin{verbatim}
    SA:  ...I'm just trying to break it down 
         as much as possible so that we can 
         find any mistakes.
\end{verbatim}  In this way, the group began to not only determined the correctness of lines of code that have been added/modified, but also began to examine the relationships \emph{between} those lines.

For example, the group began by confirming the correctness of the form of one such line of code:  \begin{verbatim}
    SA:  Final momentum equals initial momentum
         plus net force times delta t.  True?
    SC:  Yeah...
    SB:  Yes.
    SA:  O.K.  That's exactly what we have 
         here.  So this is not the problem.
         This is right.
    SD:  Yeah.
\end{verbatim}  That is, Student A \emph{(i)} read aloud and wrote down the line of code $\vec{p}_{f}=\vec{p}_{i}+\vec{F}_{\rm net}\Delta t$ while the entire group confirmed on its correct form.  This written line was then boxed, and was shortly followed up \emph{(ii)} with a similar confirmation of the line $\vec{r}_{f}=\vec{r}_{i}+\vec{v}\Delta t$ that immediately prompted \emph{(iii)} the confirmation of $\vec{v}=\vec{p}/m$.  Thus, not only do we see the group determining the correctness of added/modified lines of code as in \emph{(i)}--\emph{(iii)}, we further see confirmation with the links \emph{between} those lines.

The confirmation of the link between the lines of code \emph{(i)} and \emph{(ii)}, representing the incremental update of position and momentum in time, respectively, was evidenced not through the mere addition of the linking equation \emph{(iii)} to the list of lines added, but further through the gestures exhibited by student A.  Pointing at \emph{(iii)}, the $\vec{v}$ in \emph{(ii)}, and the $\vec{p}_{f}$ in \emph{(i)}, demonstrated that the group understood that without this linking equation \emph{(iii)}, the velocity used in \emph{(ii)} would not reflect the time updated velocity by means of \emph{(i)}.

The group ran through these types of confirmations with FPPs rapidly over the span of a few minutes.  Once the group had confirmed all the added/modified lines of code to their satisfaction, the discussion quieted down.  The FPPs were winnowed from the discussion, and after a little more off-topic discussion we find them seeking help from the instructor:  \begin{verbatim}
    SD:  Maybe we should just stare at him 
         until he comes help us...
\end{verbatim} Suddenly, a haphazard change to the code:  \begin{verbatim}
    SA:  You know what, I'm gonna try 
         something.
\end{verbatim} where Student A changed the order of magnitude of the initial momentum a few times.  This modification eventually resulted in a simulation that produced the correct output.

\subsection{Bug resolution}

At about $\SI{65}{\minute}$ into the problem, Student A changed the order of magnitude of the momentum one final time, which produced something \emph{closer} to the output that they expected:  \begin{verbatim}
    SA:  Oh wait... Oh god...
    SD:  Is it working?
\end{verbatim}  The satellite now elliptically orbited the Earth.  This marks the end of the debugging phase and the beginning of the resolution phase -- the bug had successfully been found and remedied.  Given that the only line of code modified to produce this change was the initial momentum, they began to rethink the problem:  \begin{verbatim}
    SD:  I think that is the issue is that 
         we don't have the initial momentum...
    SA:  ...momentum correct?
\end{verbatim}  That is to say, the group pursued the issue of determining the correct initial momentum with the added insight gained through debugging fundamentally correct VPython code.

\section{Discussion}

To summarize, in analyzing this particular group, we first identified the two phases in time when the group had recognized and resolved a physics bug.  We then necessarily identified the phase in-between as the process of physics debugging in ${\rm P}^{3}$, where the fundamentally correct code was taken from producing unexpected output to producing expected output.  Given our assumption that the process of computer science debugging encourages desirable strategies, we then began to analyze this process of physics debugging further for strategies unique to ${\rm P}^{3}$.

\begin{figure}
\resizebox{0.9\columnwidth}{!}{\includegraphics{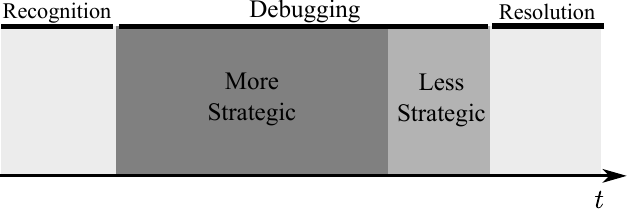}}
\caption{Physics debugging phase consisting of more and less strategic strategies specific to our case study, bounded on either side by bug recognition and bug resolution.}
\label{strategy}
\end{figure}

Given the actions exhibited during the debugging phase, we can separate them into two distinct parts: a more strategic part and a less strategic part, as shown in Fig.~\ref{strategy}.  The group initially gave indication that they were working in a considerate, thorough, and consistent manner, which we classify as more strategic.  This is contrasted by the later indications of more haphazard actions, which we classify as less strategic.  These are the two physics debugging strategies that, together, led to the resolution of the bug in this context.

The more strategic strategy was exhibited through the confirmation of individual FPPs as well as their relation to others.  Not only did the group confirm through discussion, they simultaneously wrote, boxed, and referenced equations in the code -- this helped to reduce the number of FPPs they needed to cognitively juggle at any given time \cite{Redish2003}.  This confirmation of FPPs through discussion presented a great learning opportunity for the entire group, where creative and conceptual differences could be jointly ironed-out.  Accordingly, we tentatively refer to this strategy as \textbf{self-consistency}.

Although the resolution of the bug might not be tied directly to this self-consistency, that does not negate the learning opportunities afforded to the group along the way.  Specifically, we saw the group double-checking every fundamental idea used and, possibly more importantly, the links between those ideas.  Being physically self-consistent in this manner is a desired strategy in ${\rm P}^{3}$.

The less strategic strategy was exhibited during the haphazard changes to the initial momentum.  These changes to the code that eventually resolved the bug, though one of the benefits of computation (i.e., the immediacy of feedback coupled with the undo function), could have been more thoughtful.  A deeper understanding of the physics or computation could have tipped the group off to the fact that the initial momentum was too small.  Again, this does not negate the learning opportunities afforded to the group through this less strategic strategy, which resembles \cite{Podolefsky2012} that of ``productive messing about.''  Accordingly, we tentatively refer to this strategy as \textbf{play}.

Both of these strategies identified here, self-consistency and play, provided learning opportunities to the group which are bolstered by the computational nature of the task.  In other words, the necessity of translating a collection of physical ideas into lines of code which must logically flow and the benefit of immediate visual display resulted in learning opportunities which might otherwise have been missed in an analytic task.  More research is needed to dissect these learning opportunities and to deepen our understanding of the strategies themselves.
\vspace*{0.2in}
\section{Conclusion}

This case study has described two strategies (one more and one less strategic) employed by a group of students in a physics course where students develop computational models using VPython while negotiating meaning of fundamental physics principles. These strategies arose through the group's process of debugging a fundamentally correct program that modeled a geostationary orbit. The additional data we have collected around students' use of computation is rich, and further research is needed to advance the depth and breadth of our understanding of the myriad of ways in which students might debug computational models in physics courses.

\begin{acknowledgments}
The authors thank the members of PERL@MSU for their useful comments and suggestions at various stages of this work.  We also thank Stuart Tessmer for his continued involvement in ${\rm P}^{3}$.  This work is supported in part by the CREATE for STEM Institute at MSU.
\end{acknowledgments}

\bibliographystyle{apsper}
\bibliography{./bib.bbl}

\end{document}